\documentclass[12pt]{emulateapj}
\usepackage{amsmath}
\usepackage{graphicx}
\usepackage{color}
\usepackage{natbib}

\shorttitle{Fine strand-like structure in the solar corona from MHD transverse oscillations}
\shortauthors{P. Antolin, T. Yokoyama, \& T. Van Doorsselaere}

\begin{document}

\title{Fine strand-like structure in the solar corona from MHD transverse oscillations}

\author{P. Antolin\altaffilmark{1}, T. Yokoyama\altaffilmark{2}, and T. Van Doorsselaere\altaffilmark{3}}
\affil{\altaffilmark{1}National Astronomical Observatory of Japan, 2-21-1 Osawa, Mitaka, Tokyo 181-8588, Japan\\
\altaffilmark{2}Department of Earth and Planetary Sciences, University of Tokyo, Bunkyo-ku, Tokyo 113-0033, Japan\\
\altaffilmark{3}Centre for mathematical Plasma Astrophysics, Department of Mathematics, KU Leuven, Celestijnenlaan 200B, bus 2400, B-3001 Leuven, Belgium}
\email{patrick.antolin@nao.ac.jp}

\begin{abstract}

Current analytical and numerical modelling suggest the existence of ubiquitous thin current sheets in the corona that could explain the observed heating requirements.  On the other hand, new high resolution observations of the corona indicate that its magnetic field may tend to organise itself in fine strand-like structures of few hundred kilometres widths. The link between small structure in models and the observed widths of strand-like structure several orders of magnitude larger is still not clear. A popular theoretical scenario is the nanoflare model, in which each strand is the product of an ensemble of heating events. Here, we suggest an alternative mechanism for strand generation. Through forward modelling of 3D MHD simulations we show that small amplitude transverse MHD waves can lead in a few periods time to strand-like structure in loops in EUV intensity images. Our model is based on previous numerical work showing that transverse MHD oscillations can lead to Kelvin-Helmholtz instabilities that deform the cross-sectional area of loops. While previous work has focused on large amplitude oscillations, here we show that the instability can occur even for low wave amplitudes for long and thin loops, matching those presently observed in the corona. We show that the vortices generated from the instability are velocity sheared regions with enhanced emissivity hosting current sheets. Strands result as a complex combination of the vortices and the line-of-sight angle, last for timescales of a period and can be observed for spatial resolutions of a tenth of loop radius.

\end{abstract}


\section{Introduction}

The fine scale organisation of the plasma and the magnetic field in the corona is a big unknown in solar physics. Analytical and numerical studies suggest the existence of ubiquitous thin current sheets in the corona that could explain the observed heating requirements \citep{Parker_1988ApJ...330..474P, Gudiksen_2005ApJ...618.1020G, Ofman_2009ApJ...694..502O, vanBallegooijen_etal_2011ApJ...736....3V, Peter_2012AA...548A...1P}.

Observations with TRACE, SDO and Hi-C have shown a large distribution of widths, with a tendency of the peak to be located at gradually lower scales of a few hundred kilometers with increasing spatial resolution \citep{Reale_2010LRSP....7....5R, Brooks_2013ApJ...772L..19B,Peter_2013AA...556A.104P}. Even higher resolution can be achieved through observations in chromospheric lines of partially ionised plasma in the corona. This is the case of siphon flows of cool material and coronal rain in loops, or prominences in more complex field topologies. Fine structure with widths down to 100~km have been reported \citep{Ofman_Wang_2008AA...482L...9O,Lin_2011SSRv..158..237L,Antolin_Rouppe_2012ApJ...745..152A}.

While several orders of magnitude separates the predicted widths of dissipation regions such as current sheets in 3D MHD models and the observed widths of strand-like structure, the two are generally believed to be closely linked through the nanoflare scenario \citep[e.g.][]{Aschwanden_2000ApJ...541.1059A}. In this one, a heating event would be localised to the magnetic field lines involved in the process, and the ensemble of such events combined with the heating and cooling timescales of coronal plasma would lead to the observed strand organisation \citep{Klimchuk_2006SoPh..234...41K}.

\setcitestyle{citesep={ or}}
An increasing number of reports of small amplitude transverse MHD waves in the corona have appeared in recent years \citep[for a review see][]{DeMoortel_Nakariakov_2012RSPTA.370.3193D}. \setcitestyle{citesep={;}}These are waves with a few km~s$^{-1}$ amplitude and periods of a few minutes, propagating or standing in magnetic field structures in the corona, observed in a wide range of wavelengths \citep{Okamoto_2007Sci...318.1577O,Tomczyk_2007Sci...317.1192T,Erdelyi_2008AA...489L..49E,VanDoorsselaere_2008ApJ...676L..73V,Jess_2009Sci...323.1582J,Lin_2011SSRv..158..237L,McIntosh_2011Natur.475..477M,Antolin_Verwichte_2011ApJ...736..121A,Tian_2012ApJ...759..144T,Hillier_2013ApJ...779L..16H}. In this letter we suggest an alternative mechanism for generation of strand-like structure in EUV lines by establishing a link with the ubiquitous small amplitude transverse MHD oscillations observed in the corona. This link is based on the fact that a loop subject to transverse oscillations suffers from relatively strong shear motions toward the edges, leading to Kelvin-Helmholtz (K-H) instabilities \citep{Heyvaerts_1983AA...117..220H,Uchimoto_1991SoPh..134..111U,Ofman_1994GeoRL..21.2259O,Terradas_2008ApJ...687L.115T}. The importance of this instability has been stressed in magnetic reconnection studies and coronal heating \citep{Karpen_1993ApJ...403..769K,Ofman_1994GeoRL..21.2259O,Fujimoto_1994JGR....99.8601F,Lapenta_2003SoPh..214..107L}. In the case of the corona, this instability has so far been directly observed on large scales in CMEs \citep{Foullon_etal_2011ApJ...729L...8F, Ofman_2011ApJ...734L..11O} and in much smaller scales in quiescent prominences \citep{Berger_etal_2010ApJ...716.1288B,Ryutova_2010SoPh..267...75R}. Here we suggest a more general existence of this instability in the corona and propose strand-like structure as one of its characteristic features.

\section{Numerical model and initial conditions}

We consider a density enhanced loop in a low-$\beta$ coronal environment, oscillating with an MHD standing kink mode. To this end we perform 3D MHD simulations including numerical resistivity with an MHD code based on the CIP-MOCCT scheme \citep{Kudoh_1999_CFD.8}. The present model is similar to that of \citet{Terradas_2008ApJ...687L.115T}. The magnetic field is uniformly set to $B_{0}\approx22.8~$G, and is in hydrostatic equilibrium by taking density and temperature ratios $\rho_{i}/\rho_{e}=3$, $T_{i}/T_{e}=1/3$, respectively, where the index $i$ ($e$) denote internal (external) values. We have also considered a loop with the same density ratio but no temperature variation across the boundary (thus allowing a slight decrease of magnetic field within the loop). The results are qualitatively the same and we therefore focus here on the uniform magnetic field model. We take $T_i=10^6~$K and the plasma-$\beta$ is 0.02. The initial density profile is set by the following formulae, where $\rho_e=10^{9} \mu m_p~$g~cm$^{-3}$ ($\mu=0.5$ and $m_p$ is the proton mass):
\begin{multline}\label{eq1}
\rho(x,y) = \rho_e+(\rho_i-\rho_e)\zeta(x,y)\\ 
\zeta(x,y) = \frac{1}{2}(1-\tanh((\sqrt(x^2+y^2)/R-1)b)).
\end{multline}
Here, $x$ and $y$ denote the coordinates in the plane perpendicular to the loop axis, and $z$ is along its axis, $R$ denotes the loop radius and $b$ sets the width of the boundary layer. The latter is of particular importance since it is here where resonant absorption takes place \citep{Ionson_1978ApJ...226..650I,Hollweg_1988JGR....93.5423H,Sakurai_1991SoPh..133..227S}. Here we show the results for $b=16$, that is, $\ell/R\approx0.4$, where $\ell$ denotes the width of the boundary layer. The length of the loop is $200~R$, and we set $R=1~$Mm. The simulation runs with as low as possible constant resistivity and viscosity leading to viscous Reynolds and Lundquist numbers on the order of $10^{4}$. Models with higher resistivity significantly affect the instability by increasing the growth time, reducing the size of the eddies and limiting the energy cascade to lower scales. Higher viscosity is more drastic: an order of magnitude larger value above numerical dissipation completely suppresses the instability by reducing the velocity shear.

The loop is subject to a perturbation mimicking a fundamental kink mode (longitudinal wavenumber $kR\approx$0.015), by imposing initially a perturbation for the $x-$velocity component, according to $v_{x}(x,y,z) = v_0 \sin(\pi z/L)\zeta(x,y)$, where $v_0$ is the initial amplitude. The corresponding kink phase speed is $c_k=\sqrt{(\rho_{i}v_{A_{i}}^2+\rho_{e}v_{A_{e}}^2)/(\rho_i+\rho_e)}\approx1574~$km~s$^{-1}$, where $v_{A_{i}}=1285$~km~s$^{-1}$ and $v_{A_{e}}=2225$~km~s$^{-1}$ denote the internal and external Alfv\'en speeds, respectively. Here we present results for low amplitudes of the initial kink (from 3 to 15~km~s$^{-1}$, values below $0.01~c_{k}$).

The numerical box is 512 $\times$ 256 $\times$ 50 points in the $x, y$ and $z$ directions respectively, where a small number of points is chosen in the $z$ direction because of the smoothness of the solution in this direction. Thanks to the symmetric properties of the kink mode only half the plane in $y$ and $z$ is modelled, and we set symmetric boundary conditions in all boundary planes except for the $x$ boundary planes, where periodic boundary conditions are imposed. In order to minimise the influence from side boundary conditions the spatial grids in $x$ and $y$ are non-uniform, with exponentially increasing values for distances beyond the maximum displacement. The maximum distance in $x$ and $y$ from the centre is $\approx$16~R. The spatial resolution at the loop's location is 0.0156~R. Higher resolution runs with double number of grid points in $x$ and $y$ were also performed, but no significant changes were found relevant to the present investigation.

Results from the numerical simulation are converted into observable quantities by calculating the synthetic emission in EUV lines. This is performed with the recently developed FoMo code\footnote{https://wiki.esat.kuleuven.be/FoMo} \citep{Antolin_VanDoorsselaere_2013AA...555A..74A}, which is based on the CHIANTI atomic database \citep{Dere_1997AAS..125..149D}. In this paper only the results involving imaging instruments are presented. 

\section{Results}
\begin{figure}
\epsscale{1.}
\plotone{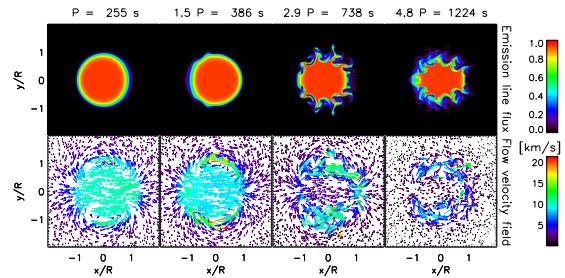}
\caption{Evolution of a cross-section at the centre of the loop. The top and bottom rows correspond to the emission line flux (for \ion{Fe}{9}~171.073~\AA), and the flow velocity field, respectively. The columns show 4 different times, written in the top of each column. See also movie 1.
\label{fig1}}
\end{figure}
\begin{figure}
\epsscale{1.}
\plotone{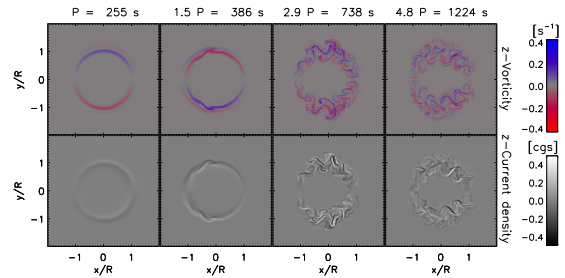}
\caption{Same as for Fig.~\ref{fig1} but for the $z$ component of the vorticity (\textit{top row}) and the $z$ component of the current density (\textit{bottom row}). See also movie 1.
\label{fig2}}
\end{figure}

\subsection{Onset of the instability}
We concentrate on the case with initial amplitude of $v_{0}=$15 km~s$^{-1}$. After the initial kink, the loop experiences a maximum displacement of $\xi_{max} /R=0.44$, and oscillates with a period $P=255$~s $\approx2L/c_{k}$. The damping time of the loop is 1047~s, within 7~\% of that predicted by resonant absorption. Fig.~\ref{fig1} and movie 1 show the time evolution of the oscillation. After a time $t\approx1.4~P$ the K-H instability sets in. As in the model of \citet{Terradas_2008ApJ...687L.115T}, the instability is generated by strong shear velocities close to the top and bottom edges ($y=\pm R$) of the perturbed flux tube. The Alfv\'en waves at the boundary increase in amplitude through resonant absorption, which leads to the generation of flows through nonlinearity (and phase mixing) and thus to the increase of the velocity shear with the external medium. Resonant absorption seems then to make the K-H instability more unstable. A linear analysis of a simplified version of our model \citep[following][]{Terradas_2008ApJ...687L.115T} leads to the instability condition:
\begin{equation}\label{KHcondition}
\frac{v_0}{v_{A_i}} > \frac{\pi}{m\sqrt{2}}\frac{R}{L}\sqrt{1+\frac{\rho_i}{\rho_e}}.
\end{equation}
Eq.~\ref{KHcondition} predicts the first unstable mode to have an azimuthal wave number of $m=3$, which grows at a rate of $0.130~P$. Fig.~\ref{fig1} shows the formation of 4 eddies around each shear layer (corresponding to $m=4$), close to the theoretical prediction. As shown by  Eq.~\ref{KHcondition} (and Eq.~1 in \citet{Terradas_2008ApJ...687L.115T}), even for low initial amplitudes, fast growth rates are ensured by small radius-to-length ratios, and should therefore be typical of long and thin loops. Modes with high wavenumbers are always unstable, but their growth is limited by dissipative effects.

\subsection{Formation of current sheets}
After a few periods the transverse oscillation has significantly damped, and most of the dynamics have turned azimuthal, concentrated around the boundary layer. This effect has been well described by resonant absorption theory in which a coupling between the kink and the Alfv\'en modes is established \citep{Goossens_2002AA...394L..39G,Goossens_2012ApJ...753..111G,Pascoe_2010ApJ...711..990P}. The resulting `Alfv\'enic' oscillations can be well seen in the vorticity maps of Fig.~\ref{fig2}. As time progresses the K-H instability significantly distorts the inhomogeneity layer, producing a widening of this layer, an apparent mixing of external and internal plasma, and spreading the Alfv\'enic oscillations to increasingly larger portions of the loop's cross-section. A main effect is the generation of large variation for the density structure in the transverse direction together with ubiquitous velocity sheared regions (see movie~1). As shown in Fig.~\ref{fig2}, these regions result in small-scale current sheets from shearing of the transverse magnetic field components. 

\begin{figure}
\begin{center}
$\begin{array}{c@{\hspace{-0.2in}}c@{\hspace{-0.2in}}c}
\includegraphics[scale=0.35]{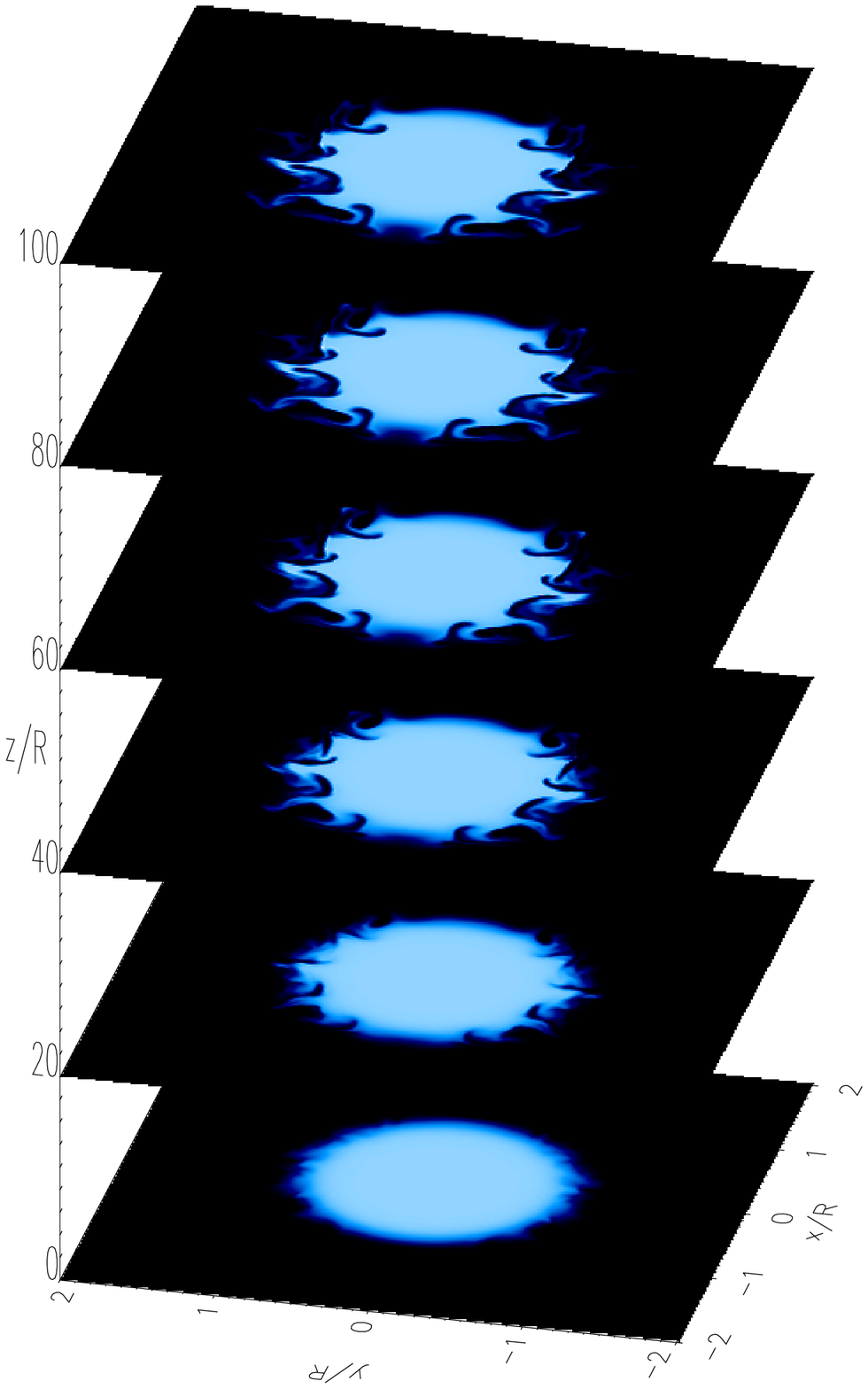}  &  
\includegraphics[scale=0.35]{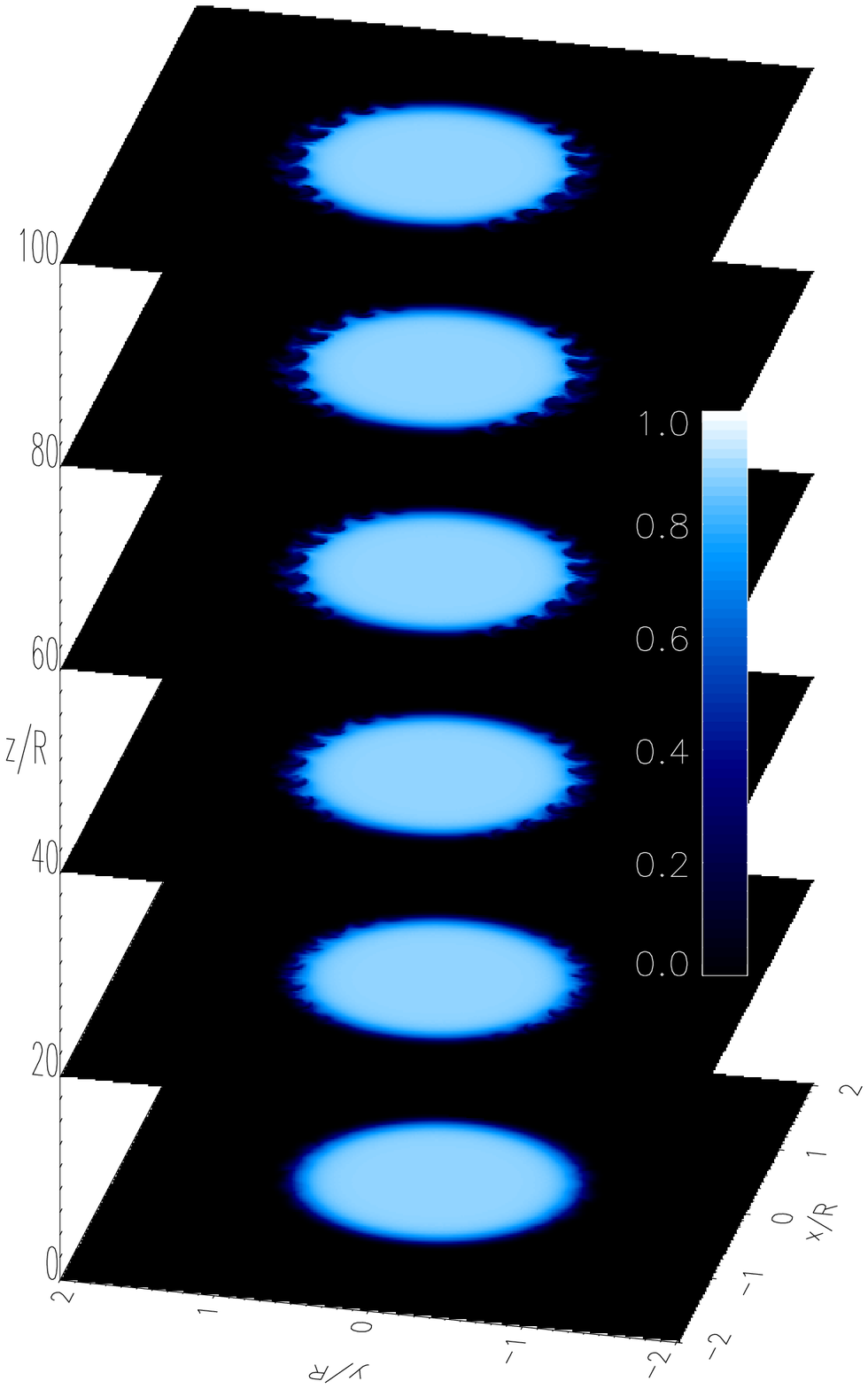} \\
\end{array}$
\caption{Snapshots of the emission line flux of \ion{Fe}{9}~171.073~\AA~for the 15~km~s$^{-1}$ (\textit{left panel}) and 3~km~s$^{-1}$ (\textit{right panel}) amplitude models, at times $t=654~$s and $t=1410~$s, respectively. Only the first half of the loop is shown. See also movies 2 and 3, in which the magnetic field lines are also shown.
\label{fig3}}
\end{center}
\end{figure}
\subsection{Strand-like structure}\label{strand}
Fig.~\ref{fig3} shows the emission line flux $\rho^2 G_{171}(T,\rho)$ in \ion{Fe}{9}~171.073~\AA~(see also movie~1), where $G_{171}(T,n)$ is the contribution function calculated assuming a coronal abundance. Movies 2 and 3 reveal that the instability develops first near the middle cross-section and a short time later is present over most of the loop, as expected from the lower velocity shear closer to the footpoints. The eddies generated by the K-H instability can be observed as distinctive emission features at the $y=\pm R$ edges of the loop. A close look at these structures for the 15~km~s$^{-1}$ reveals bright and thin layers with a 12~\% emission line flux enhancement, due to a density increase of roughly 3~\%. However, the effect on the emerging intensity of these bright structures is negligible due to their very small size.

\begin{figure}
\epsscale{1.}
\plotone{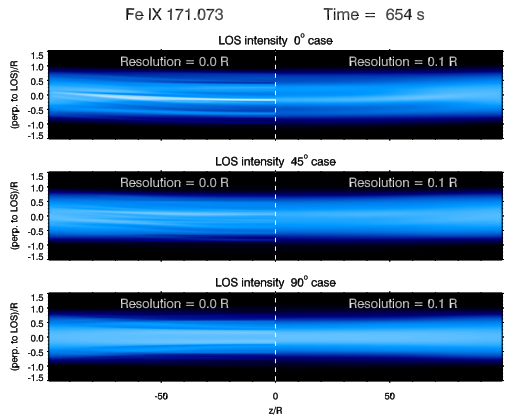}
\caption{Snapshot of intensity in \ion{Fe}{9}~171.073~\AA~for the 15~km~s$^{-1}$ amplitude model for 3 different line-of-sights: 0$^{\circ}$, 45$^{\circ}$, and 90$^{\circ}$, where the line-of-sight ray is in a plane perpendicular to the loop axis and where 90$^{\circ}$ corresponds to the oscillation direction. The left half of the loop shows the full numerical resolution, while the right half corresponds to a spatial resolution of $0.1~R$. See also movies 4 and 5.
\label{fig4}}
\end{figure}

In Fig.~\ref{fig4} and movies 4 and 5 we show the \ion{Fe}{9}~171.073~\AA~intensity, obtained by integrating the emission line flux along specific line-of-sights. For better comparison with observations we show the effect of spatial resolution by applying a Gaussian filter around each intensity pixel with a FWHM equal to $0.1~R$. The figure shows that the eddies generated by the K-H instability result in fine strand-like structure along most of the loop, for any line-of-sight. In general there is no 1-to-1 correspondence between what appears as a `strand' and an eddy, and different sizes of strands are obtained due to line-of-sight effects, with widths from 0.01~R to 0.5~R. Smaller amplitudes, as in the 3~km~s$^{-1}$ model, generate a higher number of eddies but smaller and with lower densities, which results in more uniform emission line flux and intensities. Therefore, a higher number of eddies does not strictly imply a higher number of `strands' with smaller widths. What appears as a strand is however clearly dependent on spatial resolution. At $0.1~R$ spatial resolution strand-like structure can still be observed in both models, but with larger widths (from 0.2~R to 0.5~R). Apparent crossing of strands can also be seen, especially towards the footpoints (see movies 4 and 5), which is again a combination of line-of-sight effects (as between diametrically opposed strands) or twisting, as between field lines within an eddy (see movies 2 and 3).

\section{Discussion and conclusions}

The present model shows that if a thin and long coronal loop has initially a monolithic uniform structure, a transverse perturbation of just a few km~s$^{-1}$ can significantly alter its cross-section, especially at the boundaries. Such perturbations lead to relatively large velocity shear at the boundary, enhanced by the resonant absorption mechanism, which generate Kelvin-Helmholtz instabilities. This nonlinear process widens the loop boundary layer. However, this widening is barely observed in the emission line flux due to its square dependency on density. Such unresolved boundary layer can lead to significant errors in seismology estimates \citep{Soler_2014ApJ...781..111S}. The instability leads to small-scale highly structured regions with azimuthal vortex motions. As in \citet{Ofman_2009ApJ...694..502O}, such vortices lead to magnetic twist and filamented current along the loop. However, while in that model the twist is applied as an initial condition, here it is entirely generated by the fundamental kink mode. The length scales of the vortices and currents are directly linked to those of the K-H instability eddies, decrease in scale with time and slowly expand from the boundary towards inner parts of the loop. As suggested by \citet{Ofman_2009ApJ...694..502O}, an interesting consequence of such currents is the heating of such loops. Similar filamentary current sheets have been found from the Rayleigh-Taylor instability in the context of flux emergence \citep{Isobe_2005Natur.434..478I}. Although the internal and external mixing of plasma obtained within the vortex layers in this study is only apparent and is ultimately set by numerical diffusion, such large-scale mixing has been shown to be possible for ions in the low-$\beta$ plasma conditions of the magnetopause \citep{Fujimoto_1994JGR....99.8601F}, and could therefore play an important role in heating and particle acceleration in the corona.

Due to line-of-sight effects, the filamentary structure generated by the eddies appears as strand-like structure in intensity images, which can be detected with a spatial resolution of $0.1~R$. The intensity images can give the apparent impression of crossing strands, especially towards the footpoints, a line-of-sight effect produced by twisting within K-H vortices and the existence of multiple vortices along the line-of-sight. The obtained transverse length scales in intensity images are a complex result combining the size of the eddies, their densities and line-of-sight effects. Therefore, the length scales are related  to the amplitude of the initial perturbation and the width of the boundary layer, and could serve as a seismological tool. The manifestation of strand-like structure in our model depends on the growth rate of the instability, which in turn depends on the amplitude of the initial perturbation and the width of the boundary layer \citep{Uchimoto_1991SoPh..134..111U}. For the  cases considered here the eddies appear after a time of 1.4~P to 3.25~P respectively (386~s to 830~s),  relatively fast with respect to diffusion timescales. Through auto-correlation analysis we have determined the lifetime of individual strands in our model to be around $0.5~P$ to $1~P$. Following the exact same strand is, however, very difficult since these fade in and out due to the continuous generation and destruction of the eddies, combined with line-of-sight effects. The lifetimes of large scale turbulent structures such as the eddies are only independent on viscosity when the viscous Reynolds number is large. We would therefore expect that eddies generated through K-H instability in the corona be longer lived than the ones obtained here. In our model the process of strand generation is maintained as long as eddies are generated. Resonant absorption transfers energy from the transverse motion into torsional Alfv\'en waves at the boundary layer, resulting in counterstreaming azimuthal flows. These flows, together with those generated nonlinearly by phase mixing and the K-H instability continuously create shear flows leading to the further generation of eddies. Therefore, the K-H instability benefits significantly from the resonance, from which it extracts flow energy. Once the transverse motion has damped it is not possible anymore to form large eddies, but the energy cascade to small scales still takes place.

Comparing these results with observations is far from trivial, since the definition of strands seems to be dependent on spatial resolution. \citet{Brooks_2012ApJ...755L..33B, Brooks_2013ApJ...772L..19B} and \citet{Peter_2013AA...556A.104P} report the existence of both, resolved and unresolved loops from AIA and Hi-C co-observations. The observed strands in Hi-C can be both short-lived or exist throughout the observing window \citetext{Brooks D., priv. comm.}. \citet{Winebarger_2013ApJ...771...21W} report tiny loop structures which fade in and out with lifetimes between 40~s and 80~s, whose thermodynamics match well what is expected from nanoflare heating. The length to radius ratio of these loops is close to 10, same range as in \citet{Terradas_2008ApJ...687L.115T}. Furthermore, these loops are 3 times denser than the one considered there. Assuming that the surrounding of those loops is equally denser our results suggest that small scale structure could be generated in those loops with an initial transverse perturbation of $0.1~v_{A_i}\approx20$~km~s$^{-1}$.

Evidence for small-scale transverse oscillations in coronal structures with amplitudes of a few km~s$^{-1}$ and periods of a few minutes has been provided through spectrometric or imaging measurements (see Introduction). In high resolution observations with Hi-C, \citet{Morton_2013AA...553L..10M} found little transverse wave energy, constraining amplitudes below Hi-C's spatial resolution and transverse velocities below 3~km~s$^{-1}$. However, the observations were limited to a very short time span of roughly 200~s. For proper causality analysis our model suggests that a long time span relative to the wave period is important. As already pointed out by previous work, our model also shows that transverse oscillations quickly damp out and most of the dynamics are azimuthal, and therefore could especially be detected with spectrometers. Furthermore, small scale perturbations leading to K-H instability induce mean transverse displacements of the loop as a whole that are smaller than the generated length-scales observed at $0.1~R$ resolution. In the cases considered here the maximum displacements are between $0.09~R$ and $0.4~R$ and therefore, Hi-C would only be able to detect the highest initial amplitudes in this range. The process of generation of eddies from the K-H instability continues for times longer than the damping time of the loop.

The present model does not consider thermal conduction, radiative losses, longitudinal density stratification or magnetic field variation. While these mechanisms are mostly important for generating variation along the field direction, they are only secondary for the transverse components, which decide the onset and growth of the K-H instability in this model. Furthermore, the timescales of these processes are longer than those considered here. The instability does generate twist in the field, which affects its growth but does not suppress it, a scenario similar to that in  \citet{Lapenta_2003SoPh..214..107L}. Flow is another missing ingredient, which, however, can considerably affect the damping of kink modes \citep{Gruszecki_2008AA...488..757G}. Its effect will be investigated in future studies. Here we have presented forward modelling of numerical results from 3D MHD simulations, focusing especially on reported strand-like structure from imaging instruments. The Alfv\'enic motions generated in our model result in a clear increase of line broadening, especially at the edges of loops. These results indicate that the instrumental requisites for spectrometers needed to detect such features are less strict than for the imaging instruments. Spectrometric signatures will be investigated in an up-coming paper.

\acknowledgments
We thank the referee, whose comments led to a significant improvement of the manuscript. P.A. thanks A. Malanushenko for help received in field line tracing. This work was partly supported by JSPS, Japan Society for the Promotion of Science, through a JSPS postdoctoral fellowship for foreign researchers. Numerical computations were carried out on Cray XC30 at the Center for Computational Astrophysics, National Astronomical Observatory of Japan.

\bibliographystyle{aa}
\bibliography{ms_apjl.bbl}

\end{document}